\documentclass[12pt]{iopart}
\usepackage{graphicx}
\usepackage{hyperref}
\usepackage{bm}
\usepackage{color}

\usepackage{iopams}  
\begin{document}

\title{Effects of hadron-quark phase transition on properties of Neutron Stars}

\author{Debashree Sen, and T.K. Jha}

\address{BITS-Pilani, KK Birla Goa Campus, NH-17B, Zuarinagar, Goa-403726, India}
\ead{p2013414@goa.bits-pilani.ac.in, tkjha@goa.bits-pilani.ac.in}
\vspace{10pt}
%\begin{indented}
\date{\today}
%\end{indented}

\begin{abstract}
We investigate the possible scenario of deconfinement of hyperon rich hadronic matter to quark matter at high densities and the resulting hybrid star (HS) properties are analyzed. In the relativistic mean-field framework, we construct the equation of state (EoS) of hadronic matter using the effective chiral model while the pure quark matter is described using the MIT Bag model. We revisit the hyperon puzzle and analyze the possibility of hadron-quark phase transition with proper choice of the bag constant. In static condition the maximum mass of the resultant HSs are in good agreement with the recent observational bounds on the same from high mass pulsars such as PSR J1614-2230 and PSR J0348+0432. On invoking the phenomenon of phase transition, the radius of canonical mass ($R_{1.4}$) and value of $R_{1.6}$ predicted by the model lie within the range prescribed from binary neutron star (BNS) merger detected by the LIGO-Virgo collaboration in 2017. The surface redshift obtained for the HSs also satisfy the constraints from pulsars RX J0720.4-3125 and 1E 1207.4-5209. It is noteworthy that unlike several other works, we add no modifications to the original form of the Bag model to satisfy these recent observational and empirical constraints on NS properties. We also discuss the rotational aspects of the HSs by calculating the properties like rotational mass, radius, energy density, moment of inertia at different angular velocities. The maximum bound on rotational frequency from the rapidly rotating pulsars like PSR B1937+21 and PSR J1748-2446ad are satisfied with the HS configuration. We also test the universality of our hybrid EoS in terms of normalized moment of inertia.
\end{abstract}

\noindent{\it Keywords}: Neutron Star; Hyperons; Hadron-Quark phase transition; Hybrid Star
%\maketitle

\section{Introduction}
\label{intro}
 Dense core of neutron stars (NS) may have rich phase structures like the hyperons, quarks and various boson-type condensates \cite{Glen,Glen5,Glen6,Glenq}. However, the discovery of massive NSs like PSR J1614-2230 ($M = (1.928 \pm~ 0.017) M_{\odot}$) \cite{Fonseca} and PSR J0348+0432 ($M = (2.01 \pm 0.04) M_{\odot}$) \cite{Ant} not only questions the presence of exotic species but also constrains the equation of state (EoS). Further recent estimates from the gravitational wave (GW170817) of binary neutron star (BNS) merger, detected by LIGO-Virgo collaboration in 2017 \cite{Abbott}, limits the possible value of canonical radius ($R_{1.4}$) within $R_{1.4} < 13.76$ km \cite{Fattoyev}, $12.00 \leq R_{1.4} \leq 13.45$ km \cite{Most}. Also from the discovery of BNS merger, it has been inferred that the upper limit on the radius of a 1.6 $M_{\odot}$ NS as $R_{1.6} \leq 13.3$~km \cite{Abbott,Fattoyev} while its lower limit is constrained to be $R_{1.6} \geq 10.68_{-0.04}^{+0.15}$~km \cite{Bauswein,Fattoyev}. Further a range of surface gravitational redshift $Z_S = (0.12 - 0.23)$ is obtained from the observations of two absorption features of the source spectrum of 1E 1207.4-5209 \cite{Sanwal} while for RX J0720.4-3125 the range is estimated to be $Z_S = 0.205_{-0.003}^{+0.006}$ \cite{Hambaryan}. It is thus quite interesting and challenging to satisfy all these observational and empirical constraints because the formation of exotic matter like hyperons softens the EoS, leading to reduction in maximum mass of NS. This phenomenon is often known as the ``hyperon puzzle''. According to literature, the main mechanisms to solve this puzzle are (i) considering effect of repulsive hyperon-hyperon interaction via exchange of vector mesons \cite{Bednarek,Weissenborn12,Oertel,Maslov} or scalar meson exchange \cite{Dalen}, (ii) inclusion of repulsive hyperonic three-body forces \cite{Vidana,Yamamoto,Takatsuka,Lonardoni}, (iii) effect of phase transition from hadronic matter to deconfined quark matter \cite{Ozel,Weissenborn11,Klahn,Bonanno,Lastowiecki,Dragoq1,Dragoq2,Bombaci 16,Zdunik,Masuda 16}, (iv) calculating NS properties with modified/ extended theories of gravity \cite{Doneva,Brax,Capozziello,Arapoglu,Astashenok,Astashenok2} and (v) inclusion of magnetic field \cite{Sotani2015}. In this work we study the effects of formation of hyperons ($\Lambda$, $\Sigma^{-,0,+}$ and $\Xi^{-,0}$) on the EoS and the NS properties using an effective chiral model \cite{TKJ,TKJ2,TKJ3}. We then investigate the possibility of hadron-quark phase transition and calculate the resultant hybrid star (HS) properties.
 
 The hyperons may be formed in neutron star matter (NSM) at densities when the neutron chemical potential surpasses the rest mass of the individual hyperons. However, the critical densities of formation of different hyperons strongly depend on their respective coupling constants. The most widely used hyperon coupling schemes are those based on the SU(6) \cite{Bednarek,Weissenborn12,Katayama,Schaffner-Bielich,Sulaksono,Bhowmick,Sahoo} and the SU(3) \cite{Miyatsu,Weissenborn11,Weissenborn13,Miyatsu2} quark model theories, which gives the vector and isovector couplings. Scalar couplings are obtained by reproducing the binding energies \cite{Glen,Glen5,Glen6,TKJ2,TKJ3,Arumugam} of different hyperon species in nuclear matter, constrained by certain hypernuclear studies \cite{Glen,Glen5,Rufa}. Hyperons were earlier introduced in our effective chiral model, considering the potential depth of only $\Lambda$ hyperon \cite{TKJ2,TKJ3}. Thus for a more realistic technique, the hyperon couplings are calculated in the present work, reproducing the binding energies of all the hyperon species ($\Lambda$~(-28 MeV), $\Sigma^{-,0,+}$ (+30 MeV) and $\Xi^{-,0}$ (-18 MeV) \cite{Schaffner-Bielich,Sulaksono,be}). We calculate the global properties of the NSs with the same effective chiral model \cite{TKJ,TKJ2,TKJ3}. Our model manifests chiral symmetry and the mass of all the baryons and the scalar and vector mesons are generated dynamically. The non-linearity in the model may also mimic effective three body forces, which may have decisive role at high densities \cite{three}. The model satisfies the nuclear matter saturation properties with a very few free parameters \cite{TKJ,TKJ2,TKJ3}.
 
 We then look for the possibility of hadron-quark phase transition in NSM. The phenomenon of deconfinement of hadronic matter to quark matter (QM), at high density or temperature is an exciting field in the domain of strong interaction and is an area of active research at facilities like RHIC and LHC. If the results from upcoming experiments like PREX-II can determine larger neutron skin thickness of $^{208}Pb$ then it will strongly support EoS with soft symmetry energy at high densities. This can in turn indicate possibility of phase transition in the dense NS cores \cite{Fattoyev}. In the present work the unpaired QM EoS is constructed with the well known MIT Bag Model \cite{Chodos,Glen}. From several analysis, it has been emphasized that the effects of perturbative corrections for the QM interactions \cite{Weissenborn11,Bombaci 16,Kapusta,Alford 05,Benhar,Nakazato,Uechi,Bombaci 17} can also be realized by changing the bag constant \cite{Miyatsu,Steiner,Prakash,Yazdizadeh,Burgio,Liu}. Further, with inclusion of the repulsive interactions in the thermodynamic quark potential \cite{Weissenborn11,Alford 05,Bombaci 17,Schramm,Fraga}, the resulting hybrid star (HS) properties also changes. The Nambu-Jona-Lasinio (NJL) model is also used by many authors to explain the pure quark phases \cite{Nambu,Klahn,Bonanno,Masuda 16,Steiner,Rehberg,Coelho,Masuda,Orsaria,Orsaria 2,Pereira,Mellinger,Buballa,Yang,Lenzi,Yasutake}. However, there are limitations on its usage to study the stability of HSs \cite{Yang,Benhar 2,Buballa 2,Logoteta 2}. However, certain modified forms of the NJL model like the 2+1 flavors NJL model has been recently remarkably successful to describe the quark matter model for the HSs \cite{CMLi1,CMLi2} with various well-known hadronic models to describe the hadronic phase of those HSs. Such HS configurations, obtained with this modified form of the NJL model, have even successfully satisfied the recent constraints from GW170817 \cite{CMLi3}.
 
 In case of HSs, the boundary between pure hadronic and quark matter is not so well defined but is supposed to be co-existing \cite{Glen,Glenq}. The transition properties like critical density for appearance of quarks, the density range over which mixed phase can extend and the EoS etc. are governed by the charge neutrality condition - the global charge neutrality condition (Gibbs construction (GC) \cite{Glen,Glenq,Orsaria,Rotondo}) or the local charge neutrality condition (Maxwell construction (MC) \cite{Schramm,Lenzi,Bhattacharya,LogBom}) between the two phases. Refs. \cite{Maruyama1,Maruyama2,Endo,Sotani2,Ke} have shown that beyond a limiting value of the surface tension ($\gtrsim$ 70 MeV/fm$^2$) the mixed phase becomes mechanically unstable and under such circumstances, MC provides a more physically justified and relevant way of describing the properties of HSs. However, the value of the surface tension at the hadron-quark interface is still unknown. Therefore the formation of stable hadron-quark mixed phase may still be possible if the surface tension value is not too high and then GC is favored with global charge conservation \cite{Glen,Glenq}. In the present work we assume that the surface tension of the interface is high enough to ensure that the transition occurs at a sharp interface and therefore we use MC to describe the properties of HSs. We also show our results with GC in static HS condition in \ref{Gibbs}.
 
 Several works like \cite{Miyatsu2,Yazdizadeh,Logoteta 2,Pereira 17} show that it is possible to meet the $2~M_{\odot}$ criterion of NS mass using MIT Bag model with modifications/techniques like density dependence of bag constant \cite{Yazdizadeh,Logoteta 2} or inclusion of effects like strong interactions \cite{Pereira 17} or one-gluon exchange \cite{Miyatsu2}. Various works also suggest the choice of a high value of bag constant and stiff hadronic EoS for the same purpose \cite{Logoteta 2}. The value of bag constant indeed plays a very decisive role in determining the properties of HSs. It is well established that increase in value of bag constant gives stiffer EoS and results in more massive HSs \cite{Logoteta 2,Bag,Li,Yudin}. However, the value of $B$ used in the literature ranges from $\approx ((100 - 300)$~MeV)$^4$ \cite{Steiner,Buballa,Novikov,Baym}. Therefore we choose moderate values of bag constant for which the hadron-quark crossover point lies within the density range relevant to that of NSs and calculate the global properties of the HS. The novel feature of the present work is that it shows that a proper choice of bag constant can explain massive and very compact HS configurations (consistent with the recent observational and empirical constraints on NS properties) in the presence of hyperons even with the simplest form of MIT Bag model. We do not involve any modification to the Bag model and unlike various works we do not consider any additional effects like perturbation \cite{Weissenborn11,Bombaci 16,Kapusta,Alford 05,Benhar,Nakazato,Uechi,Bombaci 17}, strong repulsive effects \cite{Pereira 17}, density dependence of bag constant \cite{Yazdizadeh,Logoteta 2} or one-gluon exchange \cite{Miyatsu2}.
  
 We also extend our analysis to study the rotational aspects of the HSs and calculate the properties like rotational mass, radius, energy density, moment of inertia at different angular velocities. As suggested recently by \cite{Breu_Rez,Lenka} it is also important to test the universality of the EoS irrespective of its composition, we study the variation of normalized moment of inertia ($I/MR^2$ and $I/M^3$) for our HSs with respect to compactness factor (M/R) in the domain of its validity of slow rotation ($P\leq 10 s$) and zero/low magnetic field ($B\leq 10^{12}$ G) \cite{Haskell}.
  
 The present manuscript is planned as follows. After discussion on the model attributes in section 1, we present the formalism employed to include hyperons in NSM, the pure quark phase and phase transition in section 2, finally culminating in the resulting HS properties and the conclusion.
 
\section{FORMALISM}
\subsection{Effective Chiral Model with baryon octet}

 The effective Lagrangian density \cite{TKJ,TKJ2,TKJ3,Sen} for the effective chiral model is given by
 
\begin{eqnarray}
\hspace*{-2.0cm}\mathcal{L} = \overline{\psi}_B \left[\left(i \gamma_{\mu} \partial^{\mu} 
- g_{\omega B}~ \gamma_{\mu} \omega^{\mu}  -\frac{1}{2} g_{\rho B}~ 
\overrightarrow{\rho_{\mu}} \cdot \overrightarrow{\tau} \gamma^{\mu} \right)-g_{\sigma B} 
\left(\sigma + i \gamma_5 \overrightarrow{\tau} \cdot \overrightarrow{\pi} \right) \right] \psi_B \nonumber \\ 
\hspace*{-1.0cm}+ \frac{1}{2} \left(\partial_{\mu} \overrightarrow{\pi} 
\cdot \partial^{\mu} \overrightarrow{\pi} + \partial_{\mu} \sigma ~ \partial^{\mu} \sigma \right)
-{\frac{\lambda}{4}} \left(x^2-x_0^2\right)^2 - \frac{\lambda B}{6} (x^2-x_0^2)^3 
- \frac{\lambda C}{8}(x^2-x_0^2)^4 \nonumber \\
-\frac{1}{4}F_{\mu\nu}F^{\mu\nu} +\frac{1}{2}\sum_B g_{\omega B}^2~x^2~\omega_\mu 
\omega^\mu -\frac{1}{4}~\overrightarrow{R_{\mu\nu}} \cdot \overrightarrow{R^{\mu\nu}}
+\frac{1}{2}~m_\rho^2 ~\overrightarrow{\rho_\mu} \cdot \overrightarrow{\rho^\mu} 
\protect\label{Lagrangian}
\end{eqnarray}

where, B and C are the scalar couplings and are evaluated at nuclear saturation density $\rho_0 = 0.153~fm^{-3}$. The subscript B denotes sum over all baryonic states viz. the nucleons and the hyperons (sumover index $B = n, p, \Lambda, \Sigma^{-,0,+}, \Xi^{-,0}$). The nucleons (N=n,p) and the hyperons (H=$\Lambda$,$\Sigma^{-,0,+}$, $\Xi^{-,0}$) interact with eachother via the scalar $\sigma$ meson, the vector $\omega$ meson (783 MeV) and the isovector $\rho$ meson (770 MeV) with $g_{\sigma_B}, g_{\omega_B}, g_{\rho_B}$ be the corresponding coupling strengths. This is a phenomenological model, applicable to the nuclear matter studies. The model embodies chiral symmetry with the scalar $\sigma$ and the pseudoscalar $\pi$ mesons as chiral partners and $x^2 = ({\pi}^2+\sigma^2)$ \cite{VolKo}. The scalar field $\sigma$ attains a vacuum expectation value (VEV) $\sigma_0$ with the spontaneous breaking of the chiral symmetry at ground state \cite{VolKo}. 

 The interaction of the scalar ($\sigma$) and the pseudoscalar ($\pi$) mesons with the isoscalar vector boson ($\omega$) was introduced to this model in \cite{Sahu2}. Such interaction generates a mass for the vector $\omega$ meson dynamically in terms of the VEV of the scalar $\sigma$ field ($\sigma_0=x_0$) via Higgs mechanism \cite{Sahu2,Sahu3}. The original chiral sigma model \cite{LeeWick} was based on the usual theory of pions which failed to produce reasonable EoS suffering from the "cusp catastrophe problem'' \cite{Boguta}. In order to overcome this problem and to yield desirable EoS consistent to the saturated nuclear matter properties, the $\omega$ meson with dynamically generated mass, was incorporated in the chiral model by \cite{Boguta,Glen86,Sahu2,Sahu3}. This phenomenological model is therefore particularly developed from the original chiral sigma model, including the dynamically generated mass of the $\omega$ meson, to obtain reasonable nuclear matter EoS in terms of saturated nuclear matter properties \cite{Sahu3,TKJ}. It was also successfully shown that the model is applicable to the high density and cold nuclear matter composition of NSs, yielding EoS compatible with the gross structural properties of NSs  \cite{Sahu3,TKJ2,TKJ3}. The model has also been successfully used to obtain the EoS of low density asymmetric matter at finite temperature \cite{Sahu}. Hence this class of phenomenological chiral models are specially developed and limited to the study of possible composition and obtain reasonable EoS of nuclear matter at different density and temperature regimes. Thus they do not find much applications in other domains of nuclear studies other than yielding reasonable EoS over a good range of density and temperature. However, considering the success of the model within its limited domain of applicability, we proceed to apply this model to obtain the hadronic part of the hybrid star matter EoS.
 
 The masses of the baryons ($m_B$) and the scalar and vector mesons can be expressed in terms of $x_0$ \cite{Sahu2,Sahu3,TKJ} as

\begin{eqnarray}
m_B = g_{\sigma_B} x_0,~~ m_{\sigma} = \sqrt{2\lambda} ~x_0,~~
m_{\omega} = g_{\omega_N} x_0
\end{eqnarray}

where, $\lambda=({m_\sigma}^2 -{m_\pi}^2)/(2{f_\pi}^2)$ is derived from chiral dynamics. $f_\pi$, being the pion decay constant, relates to the vacuum expectation value of $\sigma$ field as $<\sigma>~=~\sigma_0~=~f_\pi$ \cite{TKJ}. In the relativistic mean field treatment, we have $< \pi >~= 0$ and the pion mass $m_{\pi}=0$. Therefore the explicit contributions of pions are neglected and in this work we consider only the non-pion condensed state of matter as in \cite{TKJ,TKJ2,TKJ3}.
 
 The isospin triplet $\rho$ meson is included to account for the asymmetric nuclear matter. Its coupling strength is obtained by fixing the symmetry energy coefficient $J = 32$ MeV at $\rho_0$, given by

\begin{eqnarray}
J = \frac{C_{\rho_N}~ k_{FN}^3}{12\pi^2} + \frac{k_{FN}^2}{6\sqrt{(k_{FN}^2 + m_N^{\star 2})}}
\end{eqnarray}

 where, $C_{\rho_N} \equiv g^2_{\rho_N}/m^2_{\rho}$ and $k_{FN}=(6\pi^2 \rho_N/{\gamma})^{1/3}$.

%\vspace{0.2cm}

 The equation of motion (at T=0) for the fields and the corresponding energy density and pressure of the many baryon system are calculated in relativistic mean field approach \cite{Glen,MulSer} as a function of total baryon density $\rho$. In terms of Fermi momenta $k_B$ of a particular baryon species B, the total baryon density is 

\begin{eqnarray} 
\rho = \sum_B \rho_B =\frac{\gamma}{2\pi^2} \sum_{B} \int^{k_B}_0 dk ~k^2  
\end{eqnarray}

The value of the spin degeneracy factor $\gamma$ is 2 for this case. 
 
 At high enough momentum (density), when the nucleon chemical potential reaches the rest mass state of
the hyperons, they start appearing in dense NSM guided by the charge neutrality and chemical potential conditions. We then take those hyperons at equal footing with the nucleons. Similarly muons also appear at the expense of the electrons. 

The chemical equilibrium conditions are given as

\begin{eqnarray} 
\mu_B=\mu_n-Q_B \mu_e \label{chem_eq}\\ 
\mu_\mu=\mu_e
\end{eqnarray}

where, $\mu_n$ and $\mu_e$ are the chemical potentials of neutron and electron, respectively and $Q_B$ is the charge of baryon.

The baryon chemical potential is given by

\begin{eqnarray} 
\mu_B=\sqrt{{k_B}^2+{m^*_B}^2} ~+~g_{\omega_B} ~\omega_0 ~ +~g_{\rho_B} I_{3B}\rho_{03}
\end{eqnarray} 

where, $I_{3B}$ are the third components of isospin of the baryons and $\omega_0$ and $\rho_{03}$ are the mean field approximate or VEVs of $\omega$ and $\rho$ fields, respectively given as

\begin{eqnarray}  
\omega_0 = \frac {\sum\limits_{B} g_{\omega_B} \rho_B}{\Bigl(\sum\limits_{B} g_{\omega_B}^2 \Bigr) x^2}
\label{vector_field}
\end{eqnarray} 
and

\begin{eqnarray} 
\rho_{03}=\sum_{B}\frac{g_{\rho_B}}{m_\rho^2}I_{3_B}\rho_B 
\label{isovector_field}
\end{eqnarray} 

 The scalar equation of motion in terms of $Y=x/x_0 = m_B^{\star}/m_B$ is given by

\begin{eqnarray} 
\hspace*{-3.5cm}\sum_B \Biggl[(1-Y^2)-\frac{B}{C_{\omega_N}}(1-Y^2)^2+\frac{C}{C_{\omega_N}^2}(1-Y^2)^3  +2\frac{C_{\sigma_B}~C_{\omega_N}}{m_B^2 ~Y^4} \frac{\Bigl(\sum\limits_{B} g_{\omega_B} \rho_B\Bigr)^2}{\sum\limits_{B} {g_{\omega_B}}^2} - 2 \sum_B \frac{~C_{\sigma_B}~\rho_{SB}}{m_B~ Y} \Biggr]=0 \nonumber \\
\label{scalar_field}
\end{eqnarray}

where, the scalar density $\rho_{SB}$ of each baryon is
 
\begin{eqnarray} 
\rho_{SB}=\frac{\gamma}{2 \pi^2} \int^{k_B}_0 dk ~k^2 \frac{m_B^*}{\sqrt{k^2 + {m^{*}_{B}}^2}}
\end{eqnarray}

 Based on the above theory, the evaluated energy density ($\varepsilon$) and pressure ($P$) are as follows

\begin{eqnarray} 
\hspace*{-3cm}\varepsilon = \frac{m_B^2}{8~C_{\sigma_B}}(1-Y^2)^2-\frac{m_B^2 B}{12~C_{\omega_N}C_{\sigma_B}}(1-Y^2)^3
+\frac{C m_B^2}{16 ~C_{\omega_N}^2~ C_{\sigma_B}}(1-Y^2)^4 +\frac{1}{2Y^2}C_{\omega_N} \frac {\Bigl(\sum\limits_{B} g_{\omega_B} \rho_B\Bigr)^2}{\sum\limits_{B} {g_{\omega_B}}^2} \nonumber \\
\hspace*{-2cm}+ \frac{1}{2}~m_\rho^2 ~\rho_{03}^2 + \frac{\gamma}{\pi^2} \sum_B \int_{0}^{k_B} k^2 \sqrt{(k^2+{m_B^*}^2)} ~dk 
+ \frac{\gamma}{2\pi^2} \sum_{\lambda= e,\mu^-} \int_{0}^{k_\lambda} k^2 \sqrt{(k^2+{m_\lambda}^2)}~ dk
\protect\label{EoS1}
\end{eqnarray}

\begin{eqnarray}         
\hspace*{-3cm}P =-\frac{m_B^2}{8~C_{\sigma_B}}(1-Y^2)^2+\frac{m_B^2 B}{12~C_{\omega_N}~C_{\sigma_B}}(1-Y^2)^3
-\frac{C~ m_B^2}{16~ C_{\omega_N}^2 C_{\sigma_B}}(1-Y^2)^4
+\frac{1}{2Y^2}~C_{\omega_N} \frac {\Bigl(\sum\limits_{B} g_{\omega_B} \rho_B\Bigr)^2}{\sum\limits_{B} {g_{\omega_B}}^2} \nonumber \\ 
\hspace*{-2cm}+ \frac{1}{2}~m_\rho^2 ~\rho_{03}^2 + \frac{\gamma}{3\pi^2} \sum_B \int_{0}^{k_B} \frac{k^4}{ \sqrt{(k^2+{m_B^*}^2)}}~ dk 
+ \frac{\gamma}{6\pi^2} \sum_{\lambda= e,\mu^-} \int_{0}^{k_\lambda} \frac{k^4}{ \sqrt{(k^2+{m_\lambda}^2)}}~ dk
\protect\label{EoS2} 
\end{eqnarray} 

 Here $C_{i_B}=(g_{i_B}/m_i)^2$ are the scaled couplings with $i = \sigma, \omega, \rho$ while $C_{\omega_N}=1/{x_0^2}$.

\subsubsection{The model parameter\\}

 The model parameter set is obtained self-consistently by fixing the properties of SNM in the relativistic mean-field analysis \cite{TKJ,TKJ2,TKJ3,param1,param2} at T $= 0$. The procedure is discussed in details in ref. \cite{TKJ}. For the present work the parameter set is chosen from ref. \cite{TKJ} (set 11 of \cite{TKJ}) and is listed in table \ref{table-1}, along with the saturation properties. 

\begin{table}[ht!]
\begin{center}
\caption{Parameters of the nuclear matter models considered for the present work (adopted from previous work \cite{TKJ}) are displayed. Listed are the saturation properties such as binding energy per nucleon $B/A$, nucleon effective mass $m^{\star}_N$/$m_N$, the symmetry energy coefficient $J$, slope parameter ($L_0$) and the nuclear matter incompressibility ($K$) defined at saturation density $\rho_0$. $C_{\sigma_N}$, $C_{\omega_N}$ and $C_{\rho_N}$ are the corresponding scalar, vector and iso-vector couplings. $B$ and $C$ are the higher order couplings of the scalar field. The scalar meson mass $m_{\sigma}$ is also displayed.}
\setlength{\tabcolsep}{15.0pt}
%{\small{
%\hline
\begin{center}
\begin{tabular}{cccccccc}
\hline
\hline
\multicolumn{1}{c}{$C_{\sigma_N}$}&
\multicolumn{1}{c}{$C_{\omega_N}$} &
\multicolumn{1}{c}{$C_{\rho_N}$} &
\multicolumn{1}{c}{$B/m^2$} &
\multicolumn{1}{c}{$C/m^4$} &
\multicolumn{1}{c}{$m_{\sigma}$}\\
\multicolumn{1}{c}{($\rm{fm^2}$)} &
\multicolumn{1}{c}{($\rm{fm^2}$)} &
\multicolumn{1}{c}{($\rm{fm^2}$)} &
\multicolumn{1}{c}{($\rm{fm^2}$)} &
\multicolumn{1}{c}{($\rm{fm^2}$)} &
\multicolumn{1}{c}{(MeV)} \\
\hline
6.772  &1.995  & 5.285 &-4.274   &0.292  &510 \\
\hline
\hline
\multicolumn{1}{c}{$m^{\star}_N$/$m_N$}&
\multicolumn{1}{c}{$K$} & 
\multicolumn{1}{c}{$B/A$} &
\multicolumn{1}{c}{$J$} &
\multicolumn{1}{c}{$L_0$} &
\multicolumn{1}{c}{$\rho_0$} \\
\multicolumn{1}{c}{} &
\multicolumn{1}{c}{(MeV)} &
\multicolumn{1}{c}{(MeV)} &
\multicolumn{1}{c}{(MeV)} &
\multicolumn{1}{c}{(MeV)} &
\multicolumn{1}{c}{($\rm{fm^{-3}}$)} \\
\hline
0.85  &303  &-16.3   &32  &87  &0.153 \\
\hline
\hline
\end{tabular}
\end{center}
%}}
\protect\label{table-1}
\end{center}
\end{table}
 
The nuclear incompressibility ($K = 303$~ MeV) and symmetry energy coefficient for the model ($J = 32$~MeV) agree well with findings from \cite{k1,k2,Stone2,j1}. However, the slope parameter for the present model ($L_0 = 87$~MeV) is a bit larger than the recent limits imposed \cite{l1} although ref. \cite{rmf1} suggests its range to be $L_0 = (25 - 115)$~MeV. With the recently obtained co-relation between the symmetry energy and tidal deformability and radius $R_{1.4}$ of NS, refs. \cite{Fattoyev,ZhenYuZhu} show that the value of $L_0$ can be $\sim$ 80 MeV, which is quite consistent with that obtained with our model. The rest of the saturation properties such as the saturation density ($\rho_0 = 0.153$~$\rm{fm^{-3}}$) and binding energy per particle ($B/A = -16.3$~MeV) for SNM, obtained with our model, have the standard values \cite{rmf1}. The EoS for both SNM and pure neutron matter (PNM) with this parameter set (shown in table \ref{table-1}) are also in good agreement with the heavy-ion collision data \cite{hic} as shown in \cite{TKJ,TKJ2,TKJ3}. However, the EoS passes through the soft band of heavy-ion collision data \cite{TKJ}. Unlike various models \cite{Katayama,Sulaksono,Sahoo,Miyatsu,Miyatsu2} that yield stiff EoS and satisfy the 2 $M_{\odot}$ criteria of NS mass even in the presence of hyperons, our model yields soft EoS at high density \cite{TKJ}. The reason can be attributed to the high value of nucleon effective mass ($m_N^*=0.85~ m_N$) yielded by our model compared to other RMF models \cite{rmf,rmf1}. Also there is dominance of vector repulsive force at high densities as shown in \cite{Sahu,TKJ} as a result of which the nucleon effective mass increases again at high densities. At very low densities, the scalar and vector potentials for SNM behave linearly with density similar to RMF models like NL3 ($m_N^* = 0.60$) \cite{Sahu}. However, the deviation in behavior of the potentials for our model becomes visible with increasing density and effective mass when the highly non-linear terms of scalar field with coefficients B \& C and the mass term of the vector field present in our model (eq. \ref{Lagrangian}) become dominant. The vector potential for NL3, on the other hand, increases linearly as it does not depend on the non-linear terms of the vector meson \cite{Sahu}. Therefore in comparison to other RMF models like NL3 we obtain soft EoS which softens more with the inclusion of hyperons.

 This model along with the same parameter set (as given in table \ref{table-1}) has also been used to study the nuclear matter properties at finite temperature \cite{Sahu} and the properties of NSs with hyperon rich matter in both static \cite{TKJ2} and rotational configurations \cite{TKJ3} considering the potential depth of only the $\Lambda$ hyperon. It is to be noted that the model parameters are constrained and related to the vacuum expectation value of the scalar field and there are very few free parameters to adjust the saturation properties \cite{TKJ,Sahu}. Thus the parameter set and the model adopted in the present work are well tested and in accordance with the recent experimental and empirical estimates of nuclear saturation properties and the heavy-ion collision data. 
 
 However, the chosen parameter set (shown in table \ref{table-1}) is not unique for our model. There are few other parameter sets (e.g. sets 13 and 14 of \cite{TKJ}) that are obtained in a similar way and are also in accordance with the bounds on different nuclear saturation properties and the heavy-ion collision data. It is seen that higher ratio of the scalar to vector couplings ($C_{\sigma_N}/C_{\omega_N}$) and increasingly negative values of B lead to higher value of effective mass and lower nuclear incompressibility and thus softer EoS \cite{TKJ} yielding low mass NSs.
 
\subsubsection{Hyperon coupling constants\\}
\label{Coupling}

 We fix the value of scalar coupling constants $x_{\sigma_H}={g_{\sigma_H}}/{g_{\sigma_N}}$ within the limit ($x_{\sigma_H} \leq 0.72$) specified by \cite{Glen,Glen5,Rufa} and calculate the vector couplings $x_{\omega_H}={g_{\omega_H}}/{g_{\omega_N}}$, reproducing the binding energies of the individual hyperon species ($(B/A)_H|_{\rho_0}$ = -28 MeV for $\Lambda$, +30 MeV for $\Sigma$ and -18 MeV for $\Xi$ \cite{Schaffner-Bielich,Sulaksono,be}) in saturated nuclear matter, using relation \ref{be} \cite{Glen,Glen5,Glen6,Arumugam,TKJ2,TKJ3} 

\begin{eqnarray}         
(B/A)_H\biggr|_{\rho_0} = x_{\omega_H} ~ g_{\omega_N} ~\omega_0 + x_{\sigma_H}~ g_{\sigma_N} ~\sigma_0
\protect\label{be}
\end{eqnarray}         

We choose $x_{\rho_H}$ = $x_{\omega_H}$ as both $\rho$ and $\omega$ mesons have almost same mass and also both give rise to short range repulsive forces.

\subsection{Pure quark phase \& hadron-quark phase transition}
 
 To describe the properties of pure quark phase composed of u, d and s quarks, we adopt the well-known MIT Bag Model \cite{Chodos,Glen,Glenq,Bhattacharya,Bag,Patra}. It is characterized by a parameter called the bag constant $B$, which ensures the overall strength of interaction between the quarks. We take u ($m_u=5$ MeV) and d quark ($m_d=7$ MeV) masses to be negligible compared to the mass of s quark ($m_s=100$ MeV). The chemical potential equilibrium is given by 

\begin{eqnarray}
\mu_d = \mu_s = \mu_u + \mu_e
\protect\label{Chem_pot_uqm}
\end{eqnarray}

The individual quark chemical potentials are obtained in terms of $\mu_n$ and $\mu_e$ are as follows

\begin{eqnarray}
\mu_u=\frac{1}{3}\mu_n - \frac{2}{3}\mu_e
\end{eqnarray}

\begin{eqnarray}
\mu_s=\mu_d=\frac{1}{3}\mu_n + \frac{1}{3}\mu_e
\end{eqnarray}

The total charge is to be conserved by following the relation
\begin{eqnarray}
Q = \sum_i q_i\rho_i =0
\protect\label{charge_neutrality_uqm}
\end{eqnarray}

where, i = u, d, s and e. $q_i$ and $\rho_i$ are individual charge and density of the particles, respectively. For simplicity we take the simplest form of the MIT Bag model \cite{Glen,Glenq,Bhattacharya,Bag,Patra} without considering perturbative corrections or repulsive effects of the strongly interacting quarks. The energy density and pressure are then given by \cite{Glen,Glenq,Kapusta}
 
\begin{eqnarray}
\varepsilon_{QM} = B+ \sum_f \frac{3}{4\pi^2} \Biggl[\mu_fk_f \bigg(\mu_f^2-\frac{1}{2}m_f^2\bigg) - \frac{1}{2}m_f^4 \ln\bigg(\frac{\mu_f+k_f}{m_f}\bigg)\Biggr] 
 \protect\label{eos_e_uqm}
\end{eqnarray}

and

%\vspace*{-0.5cm}

\begin{eqnarray}
P_{QM} = -B+ \sum_f \frac{1}{4\pi^2} \Biggl[\mu_fk_f\bigg(\mu_f^2-\frac{5}{2}m_f^2\bigg)
+ \frac{3}{2}m_f^4 \ln\bigg(\frac{\mu_f+k_f}{m_f}\bigg)\Biggr] 
\protect\label{eos_P_uqm}
\end{eqnarray}

\begin{eqnarray}
{\rm{where,}}~~\mu_f=(k_f^2 + m_f^2)^{\frac{1}{2}}
\end{eqnarray}

and the total density is

\begin{eqnarray}
\rho=\sum_f \frac{k_f^3}{3\pi^2}
 \protect\label{density_upq}
\end{eqnarray}
where, f = u, d and s are the quark flavors. 

 Note that we also do not include any modification to the Bag model like density dependence of bag constant \cite{Yazdizadeh,Logoteta 2} or inclusion of effects like strong repulsive interactions \cite{Pereira 17} or one-gluon exchange \cite{Miyatsu2} etc.

 The hadronic and quark phases can also co-exist and the density range over which it can extend is determined by the local/global charge neutrality condition. In case of Maxwell construction \cite{Glen,Schramm,Bhattacharya,LogBom} $\mu_B$ is continuous while there is jump in $\mu_e$ at the interface between the two phases. The pressure remains constant in the density interval of phase transition in case of MC unlike that of GC. Therefore with MC, the pressure at the transition is given by 
 
\begin{eqnarray} 
P_{MP} = P_H (\mu_B , \mu_e) = P_Q (\mu_B , \mu_e) 
\end{eqnarray}  

while the chemical potential equilibrium is given by

\begin{eqnarray}
\mu_B^H = \mu_B^Q
\protect\label{mub_mp}
\end{eqnarray}
 
The charge neutrality conditions in MC, called the local charge neutrality condition states that unlike GC the individual hadron and quark phase must be charge neutral. It is given as

%\vspace*{-0.2cm}
\begin{eqnarray}
q_H(\mu_B,\mu_e) = 0 ~;~ q_Q(\mu_B,\mu_e) = 0
\protect\label{charge_neutrality_mpMC}
\end{eqnarray} 

\vspace{1cm}
 The crust part of the NS has a much low density is taken into account by using the BPS EoS \cite{BPS} along with the obtained hybrid EoS for the core. 

\subsection{Neutron Star Structure \& Properties}
The equations for the structure of a general relativistic spherical and static star composed of a perfect fluid were derived from Einstein's equations by Tolman \cite{tov}, Oppenheimer and Volkoff \cite{tov1}, which are

\begin{eqnarray}
\frac{dP}{dr}=-\frac{G}{r}\frac{\left[\varepsilon+P\right ]
\left[M+4\pi r^3 P\right ]}{(r-2 GM)},
\label{tov1}
\end{eqnarray}

\begin{eqnarray}
\frac{dM}{dr}= 4\pi r^2 \varepsilon,
\label{tov2}
\end{eqnarray}

with $G$ as the gravitational constant and $M(r)$ as the enclosed gravitational mass for a given choice of central energy density $(\varepsilon_c)$ and specified EoS. We have used $c=1$. The value of $r~(=R)$, where the pressure vanishes defines the surface of the star. The baryonic mass $M_B(r)$ is also calculated which is defined as

\begin{eqnarray} 
M_B(r)=\int_{0}^{R} 4\pi r^2 ~\varepsilon~ m_B \left(1 - \frac{2GM}{r}\right)^{1/2} dr
\label{barmass}
\end{eqnarray}

where, $m_B$ is the mass of baryon.
 
 Another important property of compact objects like NSs is the surface gravitational redshift given by
 
%\vspace*{-0.5cm} 
 
\begin{eqnarray}
Z_S = \Biggl(1 - \frac{2GM}{R}\Biggr)^{1/2} - 1
\label{Z}
\end{eqnarray}

 In order to account for the rotational aspects of NSs, we calculate rotational quantities like the central energy density ($\varepsilon_c$), gravitational mass ($M$), angular velocity ($\Omega$), rotational frequency ($\nu$) and moment of inertia ($I$) using the rotating neutron star (RNS) code \cite{RNS}. The limiting frequency of rotation is given by the Kepler frequency ($\nu_K$), which signifies the balance between centrifugal force and gravity.

\section{Result and Discussions}
\subsection{Neutron Star properties with pure hadronic matter including hyperons}

\begin{figure}[!ht]
\centering
\includegraphics[scale=1.1]{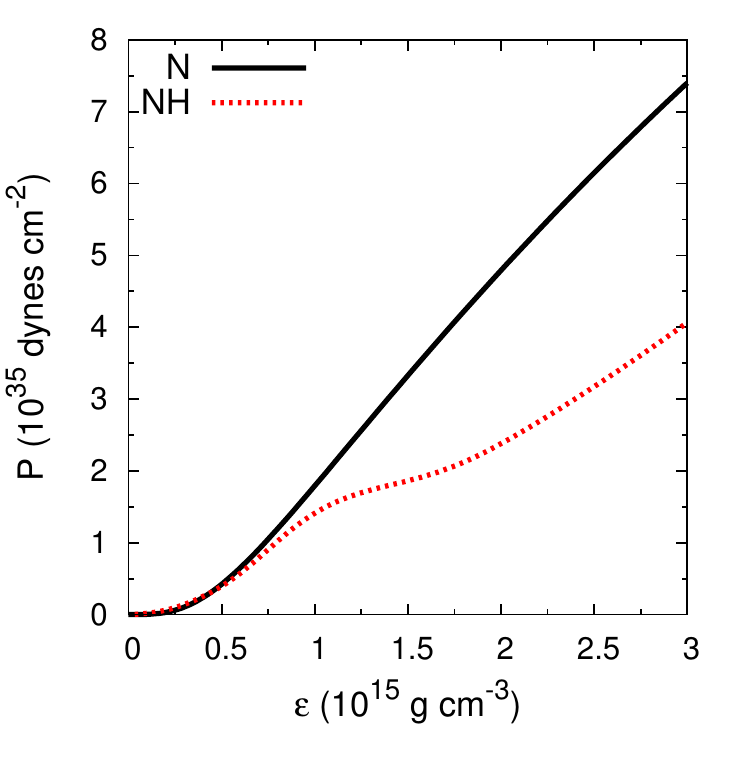}
\caption{\label{eosNH} Equation of State ($\varepsilon ~vs.~ P$) for neutron star matter with (NH) and without (N) including hyperons.}
\end{figure}

 We fix the scalar coupling of the hyperons $x_{\sigma_H}=0.65$, which is within the bound on $x_{\sigma_H}$ \cite{Glen,Glen5,Rufa} and calculate the corresponding values of $x_{\omega_H}$ according to eq. \ref{be} as discussed in section \ref{Coupling}. The corresponding EoS of hyperon rich matter (NH) is shown in fig. \ref{eosNH} and compared with the pure nucleon matter (N) (no-hyperon case). As seen from the figure, the EoS softens quite a lot due to formation of hyperons. Therefore the resultant reduction in maximum mass of NS can be well anticipated and will be discussed subsequently.
 
\begin{figure}[!ht]
\centering
\includegraphics[scale=0.7]{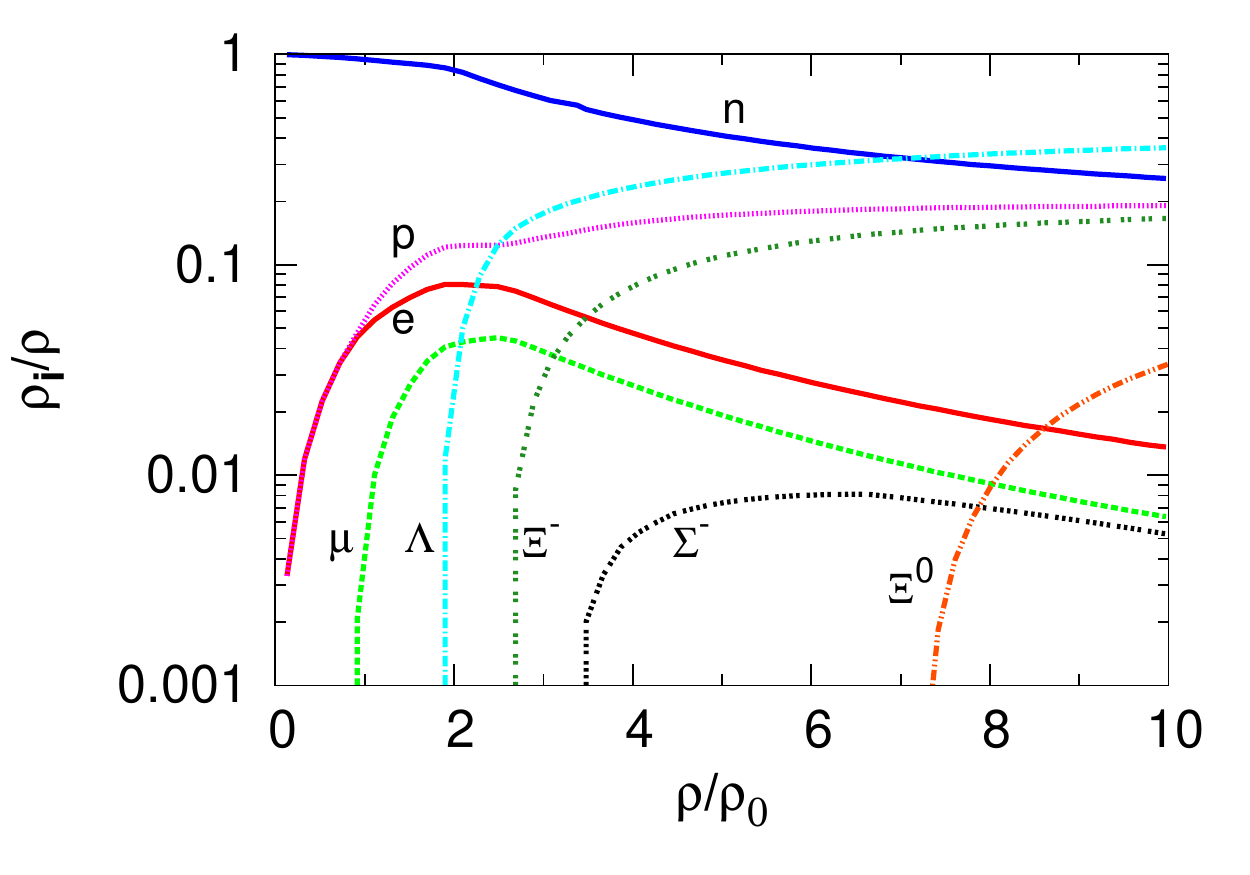}
\caption{\label{pfH} Relative particle fraction in neutron star matter including hyperons.}
\end{figure}

 In fig. \ref{pfH} we show the relative population fraction (${\rho_i}/{\rho}$) of different baryons and leptons as a function of normalized baryon density (${\rho}/{\rho_0}$). 

 Substantial amount of hyperons are formed in NSM at the expense of neutrons. $\Lambda$ appears first (approx. at 1.9$\rho_0$) and populates most in NSM compared to other hyperons. The next to form is $\Xi^-$ at density 2.6$\rho_0$, followed by $\Sigma^-$ at 3.4$\rho_0$ and $\Xi^0$ at 7.2$\rho_0$. Consistent with the findings of several works like \cite{Glen6}, we find that the formation of $\Sigma$ hyperons is quite suppressed in NSM because of their repulsive potential depth in nuclear matter. $\Sigma^-$, though formed after $\Xi^-$, unlike other hyperons its population density decreases after a certain value of density (6.7$\rho_0$). This is because of the dominance of vector repulsion on the $\Sigma$ hyperons at high densities. It is noteworthy from fig. \ref{pfH} that to balance the diminution of the negatively charged $\Sigma^-$ and maintain overall charge neutrality, the population of negatively charged $\Xi^-$ is quite high in NSM. 

\subsection{Hybrid Star properties with Maxwell construction}
 
 We then investigate the possibility of phase transition of hadronic matter to deconfined quark matter composed of u, d and s quarks. To account for the phase transition, we adopt Maxwell construction (MC) technique \cite{Schramm,Lenzi,Bhattacharya,LogBom} assuming high enough surface tension at the hadron-quark phase interface. It is suggested by \cite{Miyatsu,Steiner,Prakash,Yazdizadeh,Burgio,Liu} that the perturbative effects can also be realized by changing the bag constant $B$. Therefore, we choose the values of bag constant as $B^{1/4}$ = 180 \& 200 MeV, consistent to available literature \cite{Steiner,Buballa,Novikov,Baym}. The value has been so chosen such that the hadron-quark crossover points lie within the relevant density range of the NSs. Such a choice constrains the bag constant within the framework of our model and helps us to obtain overall stiffer EoS without involving any modification or inclusion of any effects to the Bag model. In light of GW170817 observation and measurement of tidal deformability $\Lambda_{1.4}$ \& radius $R_{1.4}$, a recent work \cite{EnPing Zhou} theoretically suggests that the values of bag constant and the repulsive interaction strength ($\alpha_4$) to be $B^{1/4}=(134.1 - 141.4)$ MeV and $\alpha_4=(0.56 - 0.91)$ for a low-spin prior while for the high-spin priors $B^{1/4}=(126.1 - 141.4)$ MeV and $\alpha_4=(0.45 - 0.91)$ considering pure quark stars. It is to be noted that our values of bag constant $B^{1/4}$ = 180 \& 200 MeV differ from \cite{EnPing Zhou} on two accounts. Firstly, \cite{EnPing Zhou} have suggested the values of $B$ and $\alpha_4$ for pure quark stars while we investigate the properties of hybrid stars in this work. Secondly, \cite{EnPing Zhou} have considered the modified form of the Bag model with repulsive interaction to obtain the properties of pure quark stars which depend not only on the value of $B$ but also $\alpha_4$. Also in context of hybrid stars, we   differ from the values of $B$ as suggested by \cite{Nandi} along with $\alpha_4$ compatible to the values of $\Lambda_{1.4}$ and radius $R_{1.4}$ obtained from GW170817 observation. One reason for the difference is because the suggested values of $B$ and $\alpha_4$ by \cite{Nandi} are obtained for hybrid stars with very few selected models like NL3, TM1 and NL3$\omega\rho$ for the hadronic part. Our effective chiral model predicts softer EoS at high density in comparison to several other similar RMF models NL3 etc \cite{TKJ,TKJ2,TKJ3} and hence needs to be complimented with stiffer EoS from the quark sector, which can be obtained with comparatively larger value of the bag constant. Moreover, \cite{Nandi} too have considered the modified form of the Bag model with repulsive interaction unlike our present work. Our choice of $B$ is in par with refs. \cite{Bhattacharya,Bag,Patra} which have also considered the MIT Bag model without repulsive interaction of quarks to account for the quark phase of the HSs.
  
\begin{figure}[!ht]
\centering
\includegraphics[scale=1.2]{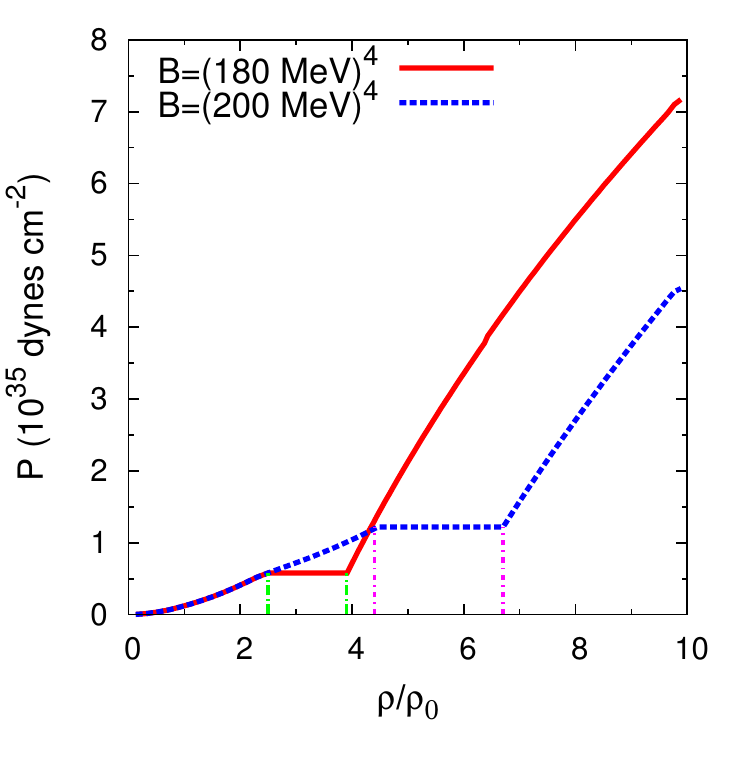}
\caption{\label{Prho} Variation of pressure with baryon density for hybrid neutron star matter including hyperons and quarks for different bag constants. The green and magenta vertical lines indicate phase transition for $B=(180$ MeV)$^4$ and $B=(200$ MeV)$^4$, respectively.}
\end{figure}

 With our choice of $B^{1/4}$ ( = 180 \& 200 MeV), consistent with \cite{Steiner,Buballa,Novikov,Baym,Bhattacharya,Bag,Patra}, we find that the transition density (chemical potential) and pressure shift to higher values with increasing bag constants. As evident from eq. \ref{eos_P_uqm}, the quark pressure decreases with increasing values of $B$ \cite{Bhattacharya,Bag}. This effect is also reflected in the EoS obtained for the hybrid star matter (HSM) with bag constants $(180~\rm{MeV})^4$ and $(200~\rm{MeV})^4$ and the density versus pressure plot (fig. \ref{Prho}) where both transition density and pressure increases with increasing bag constant.

As expected MC yields constant pressure over phase transition region unlike GC. For lower value of bag constant $(180~\rm{MeV})^4$ the hadronic phase ends at 2.5$\rho_0$ and pure quark phase initiates at 3.9$\rho_0$ while for higher bag constant $(200~\rm{MeV})^4$ the hadronic phase exists upto 4.4$\rho_0$ and the pure quark phase begins at 6.7$\rho_0$. Therefore the increase in bag constant leads to delayed phase transition and softer pure quark phase as seen in fig. \ref{Prho}. This result is consistent with that of \cite{Schramm,Lenzi,Bhattacharya,LogBom}. The formation and concentration of hyperons (mainly $\Xi^0$) are quite suppressed by the formation of quarks. Among the three, the down quark populates HSM the most, followed by up quark and then the strange quark, as dictated by their respective charge neutrality conditions and mass.

\begin{figure}[!ht]
\centering
\includegraphics[scale=1.0]{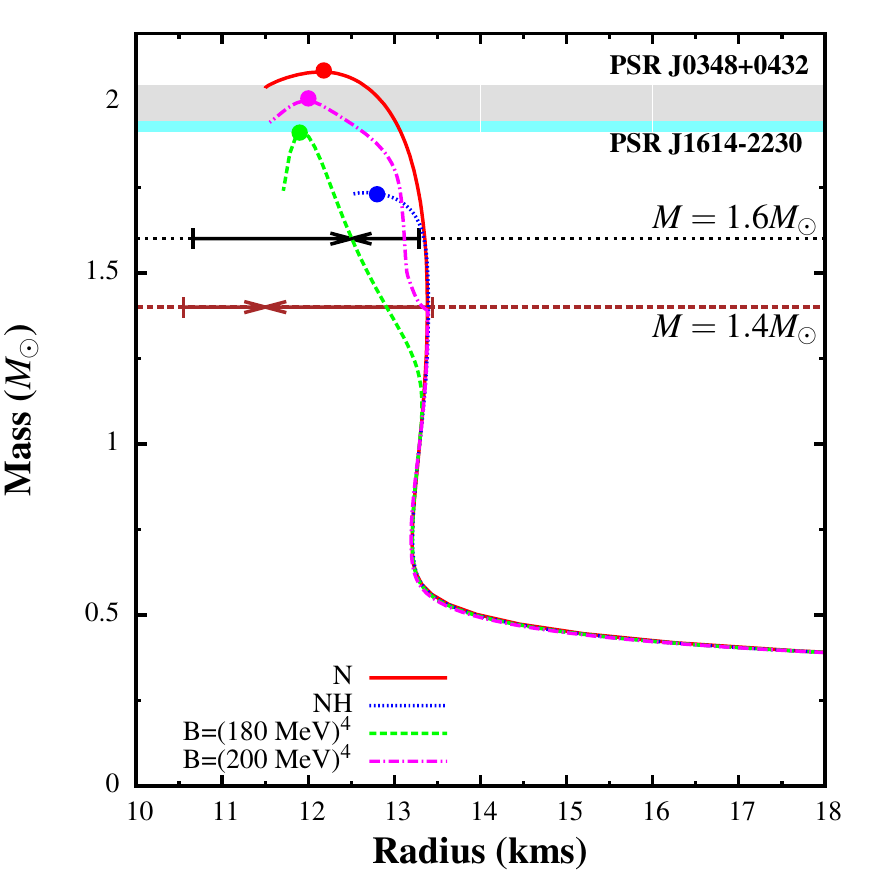}
\caption{\label{mr_all} Mass-Radius relationship in static conditions for pure nucleon star (N), neutron star with hyperons (NH) and hybrid neutron star (HS) with hyperons and quarks for two different bag constants. The points of maximum mass obtained are shown. Observational limits imposed from high mass stars PSR J1614-2230 (M = (1.928 $\pm~ 0.017) M_{\odot}$) \cite{Fonseca} and PSR J0348+0432 ($M = 2.01 \pm 0.04 M_{\odot}$) \cite{Ant} are also indicated. The brown horizontal line indicates the canonical mass ($M = 1.4~M_{\odot}$) while mass $M = 1.6~M_{\odot}$ is marked with black. Range of $R_{1.4}$ is marked according to \cite{Abbott,Fattoyev} while that of $R_{1.6}$ is marked according to \cite{Abbott} and \cite{Bauswein}.}
\end{figure}

 The properties of pure hadronic stars (with and without hyperons) and hybrid stars (HSs) like central density ($\varepsilon_c$), gravitational mass ($M$), baryonic mass ($M_B$) and radius ($R$) are calculated in static and spherical configurations. For the pure nucleon case (N) the gravitational mass is 2.10 $M_{\odot}$ with corresponding radius 12.2 km. As a consequence of softening of the EoS due to inclusion of hyperons (NH) (fig. \ref{eosNH}), we find that the gravitational mass reduces considerably to 1.73 $M_{\odot}$. With hybrid star configuration, for $B=(180$ MeV)$^4$, the maximum gravitational mass is obtained as 1.91 $M_{\odot}$ at central density 2.48 $\times 10^{15}$ g cm$^{-3}$ with corresponding radius 11.9 km. For $B=(200$ MeV)$^4$ the maximum gravitational mass is 2.01 $M_{\odot}$ at central density 1.87 $\times 10^{15}$ g cm$^{-3}$ with corresponding radius 12.1 km. The maximum baryonic mass in each case is found to be 2.26 $M_{\odot}$ and 2.34 $M_{\odot}$, respectively. The values of $R_{1.4}$ and $R_{1.6}$ obtained for $B=(180$ MeV)$^4$ are 12.8 km and 12.5 km, respectively while for $B=(200$ MeV)$^4$ they are 13.3 km and 13.1 km, respectively. In fig. \ref{mr_all} we compare the variation of gravitational mass with radius for pure nucleon star (N), NS with hyperons (NH) and hybrid NS with hyperons and quarks (HS) for two different bag constants.
 
 The various static properties of NS with different forms of matter are tabulated in table \ref{Stat.Prop}. 
 
\begin{table}[ht!]
\caption{Static neutron star properties for pure nucleon matter (N), neutron star matter with hyperons (NH) and hybrid star (HS) matter for two different bag constants ($B$). The results from hydrostatic equilibrium conditions such as the central density of the star $\varepsilon_c$ ($\times 10^{15}$ g cm$^{-3}$), the maximum gravitational mass $M$ ($M_{\odot}$), maximum baryonic mass $M_B$ ($M_{\odot}$) radius $R$ (km), $R_{1.4}$ (km) and $R_{1.6}$ (km) are displayed.}
%{\small{
%\hline
\begin{center}
\begin{tabular}{cccccccccc}
\hline
\hline
\multicolumn{1}{c}{}&
\multicolumn{1}{c}{$B^{1/4}$}&
\multicolumn{1}{c}{$\varepsilon_c$}&
\multicolumn{1}{c}{$M$}&
\multicolumn{1}{c}{$M_{B}$} &
\multicolumn{1}{c}{$R$} &
\multicolumn{1}{c}{$R_{1.4}$} &
\multicolumn{1}{c}{$R_{1.6}$} & \\
%\multicolumn{1}{c}{$M_{B}$} & 
%\multicolumn{1}{c}{$R$} & \\
%
\multicolumn{1}{c}{} &
\multicolumn{1}{c}{(MeV)} &
\multicolumn{1}{c}{($\times 10^{15}$ g cm$^{-3}$)} &
%\multicolumn{1}{c}{} &
%\multicolumn{1}{c}{} &
%\multicolumn{1}{c}{} &
%\multicolumn{1}{c}{$(\rho_0)$} &
\multicolumn{1}{c}{($M_{\odot}$)} &
\multicolumn{1}{c}{($M_{\odot}$)} &
\multicolumn{1}{c}{($km$)} &
\multicolumn{1}{c}{($km$)} &
\multicolumn{1}{c}{($km$)} & \\
\hline
N  &-   &1.22 &2.10 &2.41 &12.2 &13.4 &13.3 \\
NH &-   &1.00 &1.73 &1.91 &12.5 &13.4 &13.3 \\
HS &180 &2.48 &1.91 &2.26 &11.9 &12.8 &12.5 \\
HS &200 &1.87 &2.01 &2.34 &12.1 &13.3 &13.1 \\

\hline
\hline
\end{tabular}
\end{center}

\protect\label{Stat.Prop}
\end{table}

 We find that like many relativistic models \cite{Glen,Glen5,Glen6,Bhowmick,be} our model with pure hadron matter, including the hyperons, alone cannot resolve the hyperon puzzle. But invoking the phenomenon of hadron-quark phase transition, we successfully resolve this puzzle with the obtained hybrid EoS. Consistent with results of works done in different models and approaches \cite{Ozel,Weissenborn11,Klahn,Bonanno,Lastowiecki,Dragoq1,Dragoq2,Bombaci 16,Bombaci 17}, we find that the phenomenon of phase transition stiffens the EoS and increases the maximum gravitational mass. In the present work the maximum mass of the star increases approximately by (10 - 16)\% when the hadronic matter undergoes phase transition to quark matter in comparison to the case when only hyperons are considered. The maximum observational limit of high mass $M = (2.01 \pm 0.04)~M_{\odot}$ \cite{Ant} is successfully satisfied only when the phase transition of hadronic matter to quark matter is considered. For HS the radius for canonical mass ($R_{1.4}$) lie within the acceptable range of $R_{1.4}$ \cite{Lattimer,Fattoyev,Most}. Also for the HSs, the obtained values of $R_{1.6}$ match very well with the recent estimates of \cite{Abbott,Bauswein} (fig. \ref{mr_all}). The effect of phase transition yield overall very compact and massive HS configurations. There is also considerable increase in central density due to formation of quarks.
 
\begin{figure}[!ht]
\centering
\includegraphics[scale=1.0]{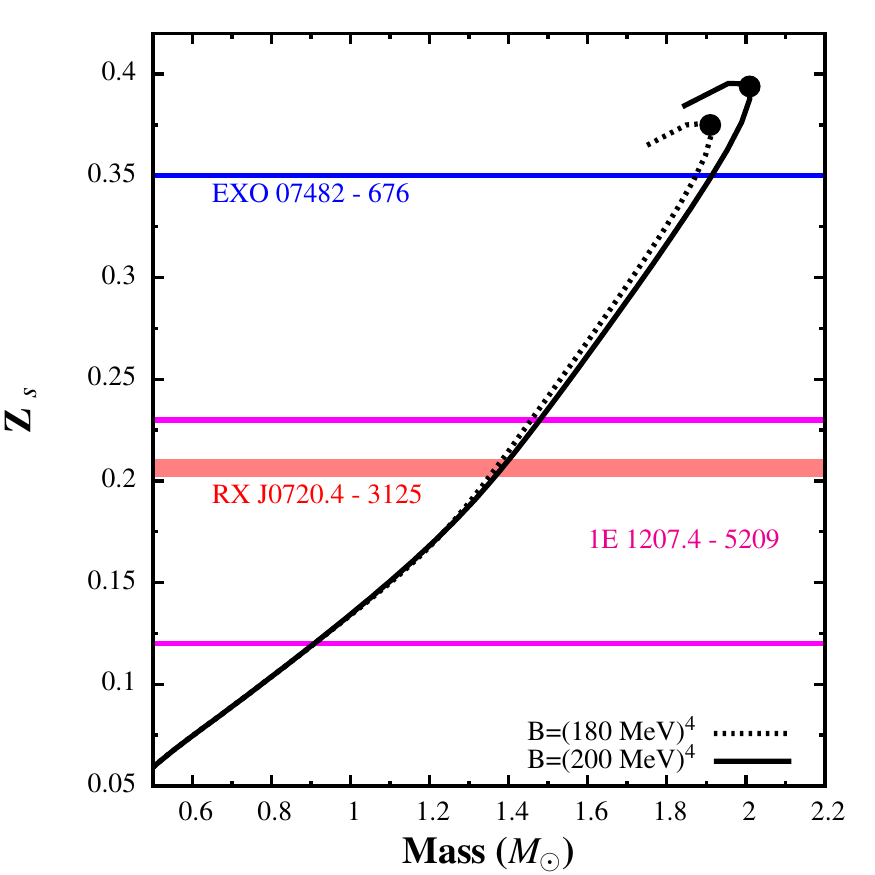}
\caption{\label{redshift} Surface gravitational redshift vs gravitational mass of hybrid neutron star with hyperons and quarks for two different bag constants. The points of maximum mass obtained are shown. Observational limits imposed on redshift from EXO 07482-676 ($Z_S = 0.35$) \cite{Cottam2002} (blue horizontal line), 1E 1207.4-5209 ($Z_S = (0.12 - 0.23)$) \cite{Sanwal} (area enclosed by magenta horizontal lines) and RX J0720.4-3125 ($Z_S = 0.205_{-0.003}^{+0.006}$ \cite{Hambaryan} (red horizontal band) are also indicated.}
\end{figure}

 In fig. \ref{redshift} we show the variation of surface gravitational redshift with mass for HS with the same values of bag constant. Consistent with the results of many works like \cite{BomLog,He,SulaksonoZ,Zhao} we find that the redshift is more in case of massive stars. The maximum redshift corresponding to maximum mass is 0.37 and 0.39 for $B=(180$ MeV)$^4$ and $B=(200$ MeV)$^4$, respectively.  
 
The obtained values of redshift with hybrid models for the specified values of $B$ lie well within the estimates of redshift from 1E 1207.4-5209 ($Z_S = (0.12 - 0.23)$) \cite{Sanwal} and RX J0720.4-3125 ($Z_S = 0.205_{-0.003}^{+0.006}$ \cite{Hambaryan} and EXO 07482-676 ($Z_S = 0.35$) \cite{Cottam2002}. The bound on redshift from EXO 07482-676 ($Z_S = 0.35$) \cite{Cottam2002} is, however, quite uncertain. The predictions of redshift from EXO 07482-676 by Cottam et al. \cite{Cottam2002} was on the basis of narrow absorption lines in the spectra of X-ray bursts from EXO 0748-676. However, in their later work \cite{Cottam2008} they obtained the lines to be much broader than that measured earlier in \cite{Cottam2002}. Also since the rotational frequency was found to be $\nu=(400 - 500)$ Hz, \cite{Lin} concluded that these spectral lines do not originate from the surface. However, more recently \cite{Baubock} suggests that the spectral lines from rotating NSs may be narrower than that predicted earlier. Thus the estimates of redshift from spectral analysis of X-ray bursts from EXO 0748-676 are still uncertain.

\begin{figure}[!ht]
\centering
\includegraphics[scale=1.0]{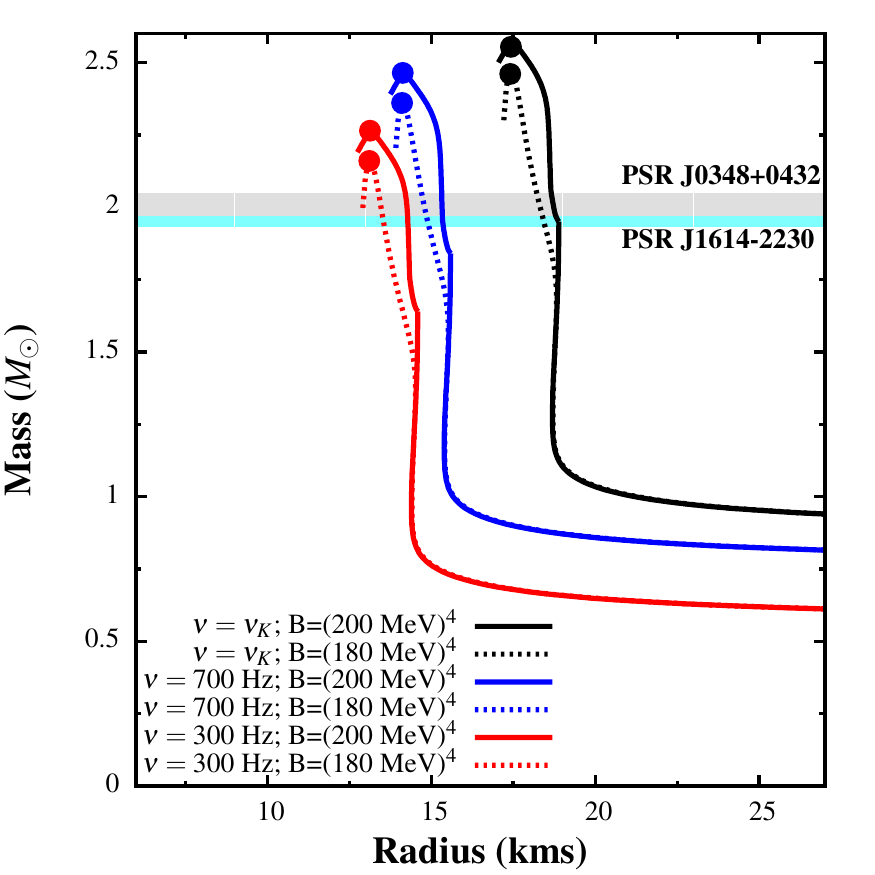}
\caption{\label{mr_rot} Mass-Radius relationship for hybrid neutron star with hyperons and quarks for two different bag constants, rotating at various frequencies. The points of maximum mass obtained are shown. Observational limits imposed from high mass stars PSR J1614-2230 (M = (1.928 $\pm~ 0.017) M_{\odot}$) \cite{Fonseca} and PSR J0348+0432 ($M = 2.01 \pm 0.04 M_{\odot}$) \cite{Ant} are also indicated.}
\end{figure}

 We now discuss the effects of rotation on the properties of HSs, calculated using the RNS code \cite{RNS}. Fig. \ref{mr_rot} shows the variation of gravitational mass with radius at different angular velocities.
 
 As expected both gravitational mass and radius increases in rotating case (fig. \ref{mr_rot}) compared to that in static case (fig. \ref{mr_all}). Also consistent with results of many works like \cite{Haensel,Cipolletta,Wei,Lenka} we find that the gravitational mass increases with rotational frequency (angular velocity) since in rotating case the centrifugal force, playing important role in determining both mass and radius of NSs, also increases with increasing frequency. In case of $\nu=300$ Hz, $B=(180$ MeV)$^4$ yields maximum mass $2.16~M_{\odot}$ at central density 2.31 $\times 10^{15}$ g cm$^{-3}$ with corresponding radius 13.1 km while $B=(200$ MeV)$^4$ gives maximum mass $2.28~M_{\odot}$ at central density 3.25 $\times 10^{15}$ g cm$^{-3}$ with corresponding radius 13.2 km. The maximum baryonic mass obtained for the respective bag constants are $2.48~M_{\odot}$ and $2.54~M_{\odot}$. For $\nu=700$ Hz, lower value of $B$ yields maximum mass $2.38~M_{\odot}$ at central density 2.87 $\times 10^{15}$ g cm$^{-3}$ with corresponding radius 14.0 km while a higher value of $B$ gives maximum mass $2.47~M_{\odot}$ at central density 3.64 $\times 10^{15}$ g cm$^{-3}$ with corresponding radius 14.1 km. The maximum baryonic mass obtained for the respective bag constants are $2.57~M_{\odot}$ and $2.93~M_{\odot}$. At Kepler frequency the maximum mass is $2.48~M_{\odot}$ at 3.36 $\times 10^{15}$ g cm$^{-3}$ with corresponding radius 17.4 km for $B=(180$ MeV)$^4$. For $B=(200$ MeV)$^4$ the maximum mass is $2.59~M_{\odot}$ at 4.13 $\times 10^{15}$ g cm$^{-3}$ with corresponding radius 17.5 km. For respective bag constants, the maximum baryonic mass are $2.96~M_{\odot}$ and $3.22~M_{\odot}$. The increase in the central density for the higher angular frequencies leads to more compact stellar structures. For example, the central density of the star increases by almost 24\% when angular frequency increases from 300 Hz to 700 Hz for the same value of the bag constant $B=(180$ MeV)$^4$.  
 
\begin{figure}[!ht]
\centering
\includegraphics[scale=1.0]{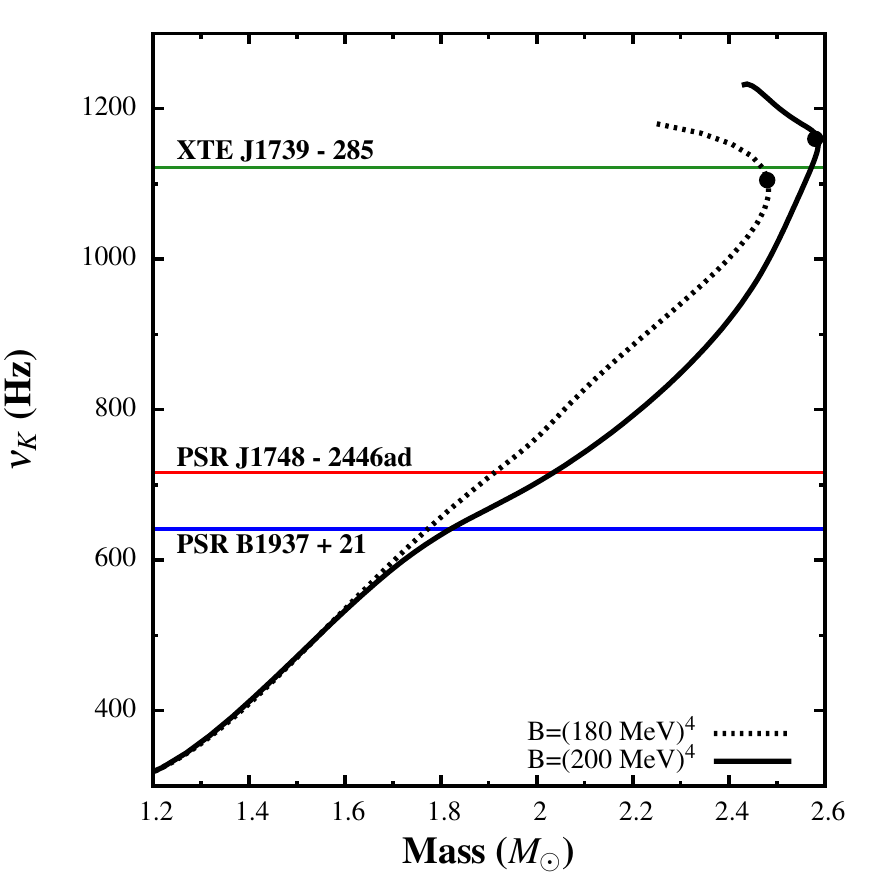}
\caption{\label{mn} Rotational frequency versus gravitational mass for hybrid neutron star with hyperons and quarks for two different bag constants, rotating at Kepler velocity. The points of maximum mass obtained are shown. Observational limits imposed on rotational frequency from rapidly rotating pulsars like PSR B1937+21 ($\nu = 633$ Hz) \cite{Backer} and PSR J1748-2446ad ($\nu = 716$ Hz) \cite{Hessels} and XTE J1739-285 ($\nu=1122$ Hz) \cite{Kaaret} are also indicated.}
\end{figure}

 Fig. \ref{mn} depicts the variation of rotational frequency of HS with respect to gravitational mass at Keplerian velocity with variation of central density.
 
 The rotational frequency profile is obtained with variation of central density at Keplerian velocity. As expected, high mass NSs can sustain fast rotations, which in this case is the star with larger value of the bag constant. For $B=(180$ MeV)$^4$ and $B=(200$ MeV)$^4$ the maximum values of $\nu_K$ are 1105 Hz and 1160 Hz, respectively. Our estimates of rotational frequency satisfy the constraints on the same from PSR B1937+21 ($\nu = 633$ Hz) \cite{Backer} and PSR J1748-2446ad ($\nu = 716$ Hz) \cite{Hessels}. For $B=(200$ MeV)$^4$ the constraint on the same from XTE J1739-285 ($\nu=1122$ Hz) \cite{Kaaret} is also satisfied. In fact \cite{Hessels,Prak} also state that
the mass of the pulsar PSR J1748-244ad to be $< 2~M_{\odot}$ rotating at 716 Hz. Our results agree with the same to a good extent for both the values of bag constant.
 
\begin{figure}[!ht]
\centering
\includegraphics[scale=0.86]{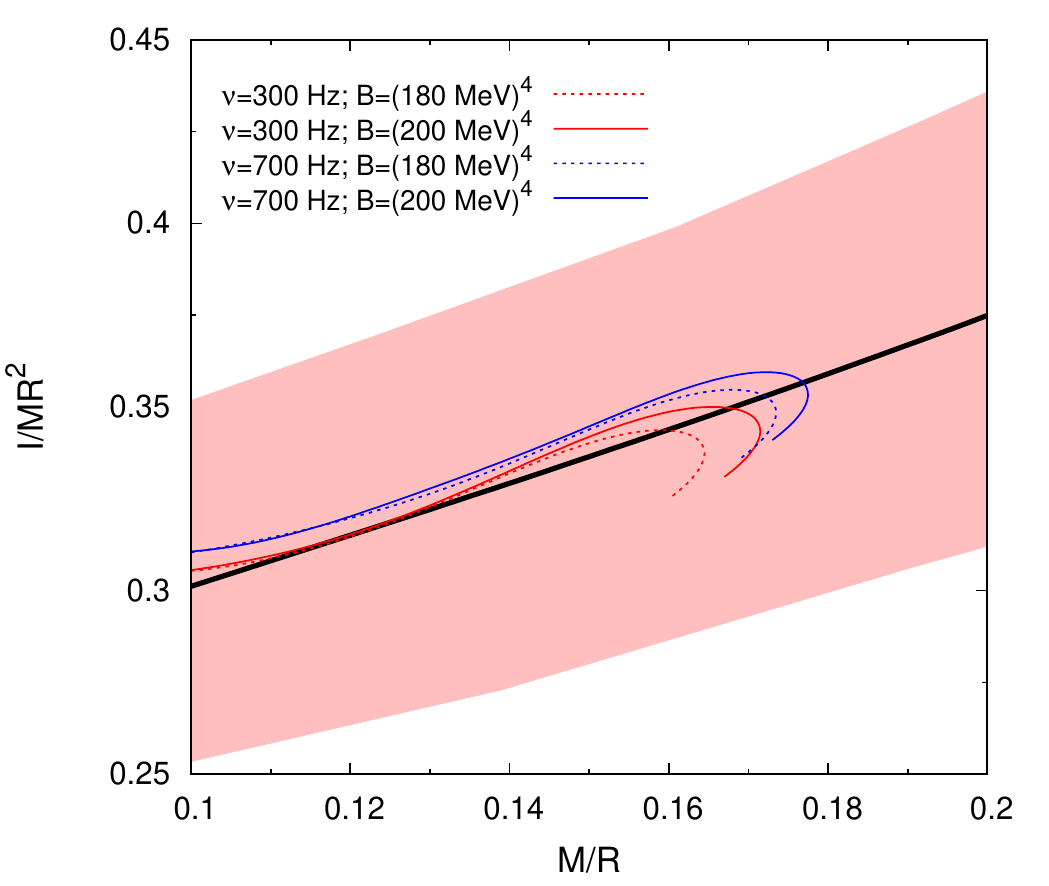}
\caption{\label{mI} Normalized moment of inertia ($I/MR^2$) versus compactness factor for hybrid neutron star with hyperons and quarks for two different bag constants, rotating at two different frequencies. The black line indicates the fitted value of normalized I in \cite{Lat_Sch} for slow rotation. The shaded pink region represents the uncertainity area of the fitted function normalized $I$ in \cite{Breu_Rez}}
\end{figure}

\begin{figure}[!ht]
\centering
\includegraphics[scale=1.0]{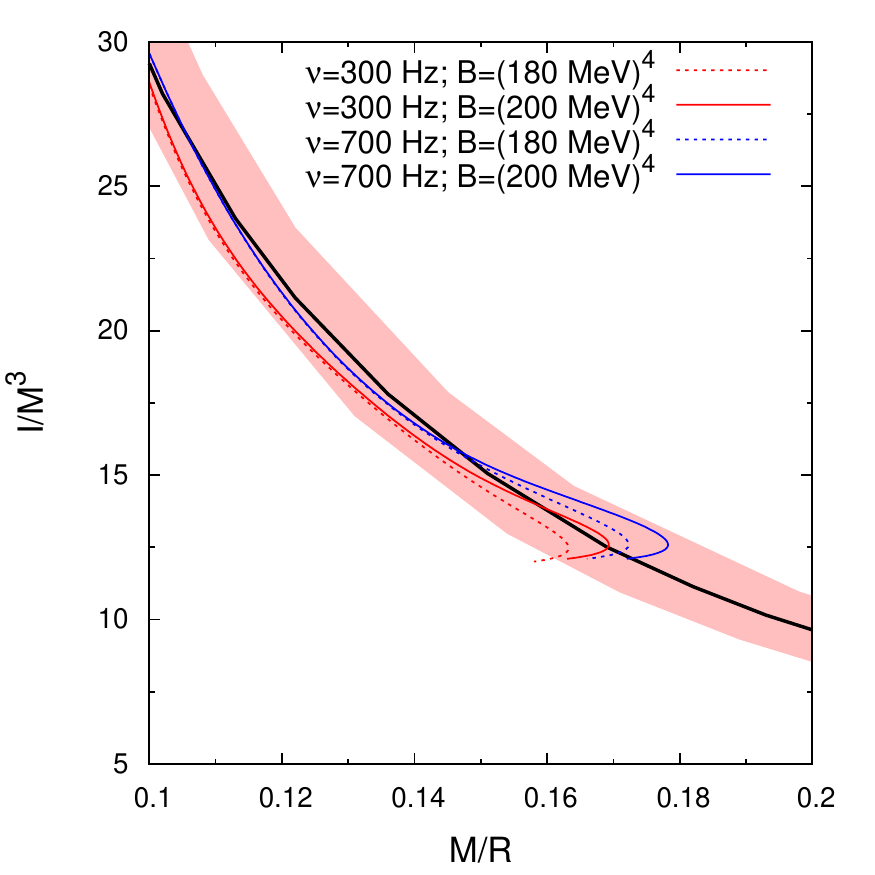}
\caption{\label{mI2} Normalized moment of inertia ($I/M^3$) versus compactness factor for hybrid neutron star with hyperons and quarks for two different bag constants, rotating at two different frequencies. The black line indicates the fitted value of normalized I in \cite{Lat_Sch} for slow rotation. The shaded pink region represents the uncertainty area of the fitted function normalized $I$ in \cite{Breu_Rez}}
\end{figure}

 Since it is imperative to test the universality of the EoS irrespective of its composition \cite{Breu_Rez,Lenka}, we study the variation of normalized moment of inertia $I/MR^2$ in fig. \ref{mI} and $I/M^3$ in fig. \ref{mI2} with respect to compactness factor ($M/R$) for HS configuration with same bag constant, rotating at frequencies $\nu=300,700$ Hz. 
 
 The moment of inertia is larger for massive NSs as they can support higher rotational frequency. It is seen from figs. \ref{mI} and \ref{mI2} for rotational frequencies $\nu=300,700$ Hz the obtained values of moment of inertia are quite consistent with the range specified for normalized value of $I$ in the slow-rotation approximation \cite{Breu_Rez,Lat_Sch}. Thus for lower values of rotational frequency (slow rotation approximation) the universality of our EoS in terms of normalized moment of inertia holds quite well.

 The various rotational properties of HS with different forms of matter are tabulated in table \ref{Rot.Prop}.

\begin{table}[ht!]
\caption{Rotational neutron star properties for hybrid star matter for two different bag constants ($B$). The results such as the central density of the star $\varepsilon_c$ ($\times 10^{15}$ g cm$^{-3}$), the maximum gravitational mass $M$ ($M_{\odot}$), maximum baryonic mass $M_B$ ($M_{\odot}$) radius $R$ (km) and maximum moment of inertia $I$ ($\times 10^{45}$ g cm$^{2}$) are displayed.}
%{\small{
%\hline
\begin{center}
\setlength{\tabcolsep}{1.2pt}
\begin{tabular}{cccccccccc}
\hline
\hline
%\multicolumn{1}{c}{}&
\multicolumn{1}{c}{$\nu$} &
\multicolumn{1}{c}{$B^{1/4}$}&
\multicolumn{1}{c}{$\varepsilon_c$}&
\multicolumn{1}{c}{$M$}&
\multicolumn{1}{c}{$M_{B}$} &
\multicolumn{1}{c}{$R$} &
%\multicolumn{1}{c}{$\nu_K$} &
\multicolumn{1}{c}{$I$} & \\
%\multicolumn{1}{c}{$M_{B}$} & 
%\multicolumn{1}{c}{$R$} & \\
%
%\multicolumn{1}{c}{} &
\multicolumn{1}{c}{($Hz$)} &
\multicolumn{1}{c}{(MeV)} &
\multicolumn{1}{c}{($\times 10^{15}$ g cm$^{-3}$)} &
%\multicolumn{1}{c}{} &
%\multicolumn{1}{c}{} &
%\multicolumn{1}{c}{} &
%\multicolumn{1}{c}{$(\rho_0)$} &
\multicolumn{1}{c}{($M_{\odot}$)} &
\multicolumn{1}{c}{($M_{\odot}$)} &
\multicolumn{1}{c}{($km$)} &
\multicolumn{1}{c}{($\times 10^{45}$ g cm$^{2}$)} & \\
\hline
300 &180 &2.31 &2.16 &2.48 &13.1 &2.53 \\
    &200 &3.25 &2.28 &2.54 &13.2 &2.76 \\
\hline    
700 &180 &2.87 &2.38 &2.57 &14.0 &3.28 \\
    &200 &3.64 &2.47 &2.93 &14.1 &3.51 \\    
\hline    
$\nu_K$ &180 &3.36 &2.48 &2.96 &17.4 &5.01 \\
        &200 &4.13 &2.59 &3.22 &17.5 &5.43 \\    
\hline
\hline
\end{tabular}
\end{center}

\protect\label{Rot.Prop}
\end{table}

 Therefore with phase transition the observational bounds on static properties like high mass ($\geq 2 M_{\odot}$) \cite{Ant} and the recent empirical estimates of $R_{1.4}$ \cite{Lattimer,Abbott,Fattoyev,Most} and $R_{1.6}$ \cite{Bauswein,Fattoyev} of NSs obtained from the GW data of BNS merger (GW170817) \cite{Abbott} are successfully satisfied. The constraints on redshift \cite{Sanwal,Hambaryan} are also successfully met with phase transition. The phenomena of hadron-quark phase transition also plays important role in determining the rotational properties and satisfying the observational constraint on rotational frequency \cite{Backer,Hessels,Kaaret} and the range of normalized moment of inertia \cite{Lat_Sch,Breu_Rez}.
 
  We show that with the mechanism of phase transition, the bounds on the aforesaid properties of HSs are not only successfully satisfied without any modifications to the basic form of MIT Bag model but also within the framework of general relativity with original the TOV eqs. \ref{tov1} and \ref{tov2} \cite{tov,tov1}. However, as discussed in the introduction section, there are also other mechanisms to solve the hyperon puzzle apart from hadron-quark phase transition. Works like \cite{Doneva,Brax,Capozziello,Arapoglu,Astashenok,Astashenok2} show that many modified/extended gravity theories, used to reformulate the TOV equations, can also successfully resolve the hyperon puzzle. Such theories of gravity \cite{Astashenok2} can throw light on important aspects of static NSs like higher order curvature effects from GW170817. Also like several works \cite{Bednarek,Weissenborn12,Oertel,Maslov}, it is possible to incorporate the strange mesons like $\phi$ and $\sigma^*$ in our hadronic model as a separate mechanism to solve the hyperon puzzle. This needs proper reformulation of the model and the EoS with re-calculated model parameters and hyperon couplings.

%\newpage
\section{Conclusion}
 We investigated the possible scenario of hadron-quark phase transition with Maxwell construction in cold dense core of massive stars containing hyperons. The composition and global properties of the stars such as the mass and radius are calculated and compared between the different prospects of NSM containing pure nucleons, hyperons and quarks. The present study concludes that recent observational bounds set by high mass pulsars like PSR J1614-2230 \cite{Fonseca} and PSR J0348+0432 \cite{Ant} can be fulfilled in static case with our model only when the hadronic matter including hyperons undergoes phase transition to deconfined quark matter with a proper choice of bag constant. Further the empirical estimates of $R_{1.4}$ and $R_{1.6}$ \cite{Lattimer,Abbott,Fattoyev,Most,Bauswein} from observations of GW170817 are also satisfied for the HS configurations. This work also highlights the results with respect to the underlying stiffness in the EoS with the phenomenon of phase transition to quark matter. It is seen that a proper choice of the bag constant, relevant to NS densities, can explain massive and very compact NS configurations even with the simplest form of MIT Bag model without involving any modification to the Bag model or additional effects like the perturbative corrections or strong repulsive interactions etc. The surface redshift obtained for the HSs also satisfy the constraints from certain pulsars like RX J0720.4-3125 \cite{Hambaryan} and 1E 1207.4-5209 \cite{Sanwal}. The rotational properties of HSs are also analyzed at different angular velocities. The maximum bound on rotational frequency from rapidly rotating pulsars like PSR B1937+21 \cite{Backer}, PSR J1748-2446ad \cite{Hessels} and XTE J1739-285 \cite{Kaaret} are satisfied with the HS configuration. The universality relation in terms of normalized moment of inertia also holds quite good for our hybrid EoS. The imposed constraints on NS properties in both static and rotational counterparts are quite well satisfied with our hybrid EoS constructed without any modification to the simplest form of MIT Bag model. In static condition of HSs, we also compare the results obtained with Gibbs construction to that obtained with Maxwell construction for HSs (\ref{Gibbs}). It will be interesting to see the results from upcoming experiments such as PREX-II, to determine the neutron skin thickness of $^{208}Pb$. Larger neutron skin thickness would support EoS with soft symmetry energy at high densities and hence can indicate phase transition in the dense core of neutron stars.

\ack
 
The authors are thankful to Dr. Kinjal Banerjee, Assistant Professor, Department of Physics, BITS-Pilani, KK Birla Goa Campus, for useful suggestions for the manuscript. 

\appendix

\section{Static properties of hybrid stars with Gibbs construction}
\label{Gibbs}

As discussed in the introduction section, the value of the surface tension at the hadron-quark interface is still unknown in existing literature and in case of low values of surface tension ($\lesssim$ 70 MeV/fm$^2$) the presence of quark matter in NSs enables the hadronic regions of the mixed phase to become more isospin symmetric than in the pure phase by transferring electric charge to the quark phase. Gibbs construction (GC) is then favored over Maxwell construction (MC) and the global charge conservation must be satisfied \cite{Glen,Glenq}. Therefore, we now investigate the static properties of HSs with GC and compare them to that obtained with MC. We use our same effective chiral model \cite{TKJ,TKJ2,TKJ3} with the same model parameter as given in table \ref{table-1} for the hadronic part of HS.

 The Gibbs criteria \cite{Glen,Glenq,Li,Logoteta 2,Orsaria,Rotondo} for the global charge neutrality condition states that the total mixed phase must be charge neutral and it is given by

\begin{eqnarray}
\chi\rho_c^Q + (1-\chi)\rho_c^H + \rho_c^l =0
\protect\label{charge_neutrality_mpGC}
\end{eqnarray}

where, $\rho_c^Q$,~$\rho_c^H$ and $\rho_c^l$ are the total charge densities of quarks, hadrons and leptons, respectively and the volume fraction for quark is given by $0\leq\chi\leq1$. Then $\chi=0~ \& ~1$ denotes pure hadronic and pure quark phases, respectively. The pressure and chemical potentials of the two phases are related by the following equations

\begin{eqnarray}
P_{MP}=P_H(\mu_B,\mu_e)=P_Q(\mu_B,\mu_e)
\protect\label{eos_P_mp}
\end{eqnarray}

and

\begin{eqnarray}
\mu_B^H = \mu_B^Q~;~~\mu_e^H = \mu_e^Q
\protect\label{mube_mp}
\end{eqnarray}

The energy density and baryon density of the mixed phase are respectively given by

%\vspace*{-0.5cm} 

\begin{eqnarray}
\varepsilon_{MP}=\chi\varepsilon_Q + (1-\chi)\varepsilon_H
\protect\label{eos_e_mp}
\end{eqnarray}

and

\begin{eqnarray}
\rho_{MP}=\chi\rho_Q + (1-\chi)\rho_H
\protect\label{eos_rho_mp}
\end{eqnarray}

\begin{figure}[!ht]
\centering
\includegraphics[scale=1.2]{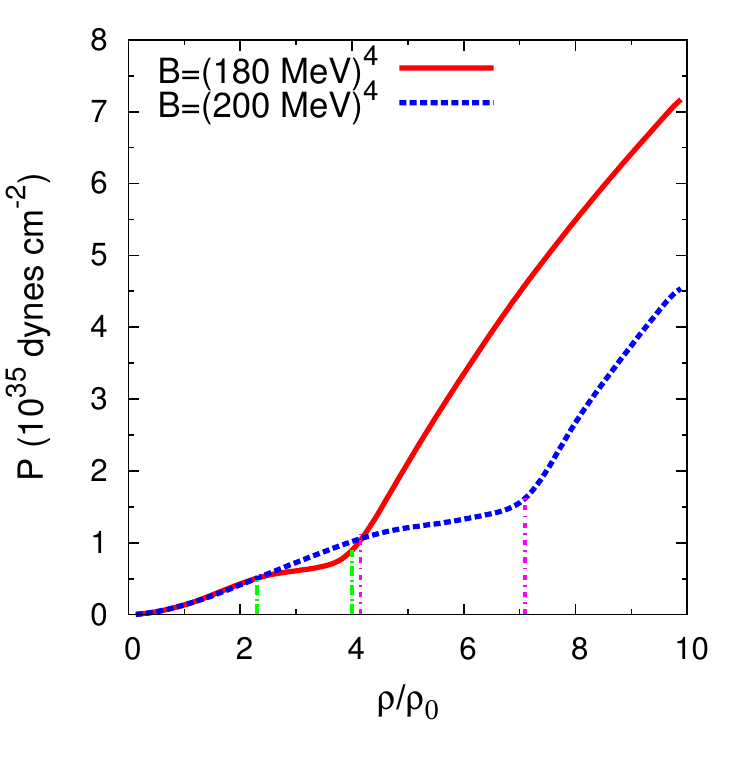}
\caption{\label{PrhoG} Variation of pressure with baryon density for hybrid neutron star matter including hyperons and quarks with Gibbs construction for different bag constants. The green and magenta vertical lines indicate the range of mixed phase for $B=(180$ MeV)$^4$ and $B=(200$ MeV)$^4$, respectively.}
\end{figure}

We obtain the hybrid EoS with GC and in fig. \ref{PrhoG} we plot the density versus pressure with the same values of Bag constant as in case of Maxwell construction.

 Comparing figs. \ref{PrhoG} and \ref{Prho}, with GC we obtain smoother transitions since in this case both the baryon and electrical chemical potentials are continuous whereas in case of MC the electrical chemical potential undergoes a sharp jump at the hadron-quark phase boundary with the baryon chemical potential remaining continuous \cite{Bhattacharya}. Unlike GC, the pressure in the region of phase transition remains constant in case of MC. Also consistent to works like \cite{Bhattacharya, Logoteta 2}, we find that in case of GC the transition occurs little early at 2.3$\rho_0$ and extends till 4.0$\rho_0$ for $B=(180$ MeV)$^4$ while for $B=(200$ MeV)$^4$ the mixed phase occurs between (4.1 - 7.1)$\rho_0$.

\begin{figure}[!ht]
\centering
\includegraphics[scale=1.0]{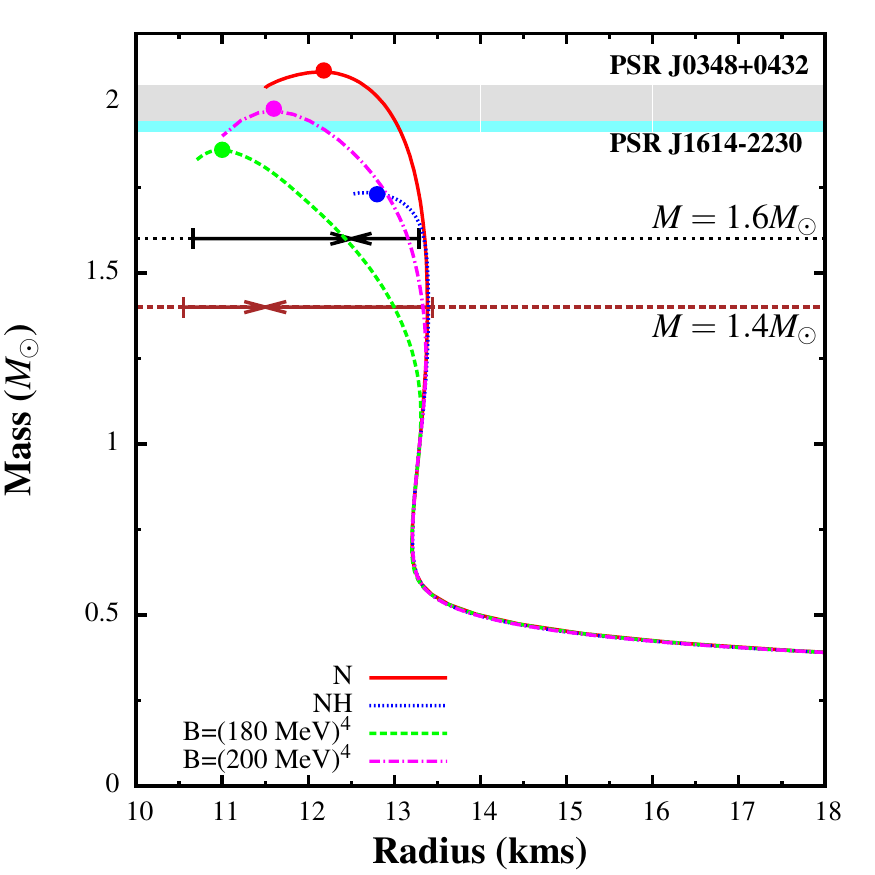}
\caption{\label{mr_allG} Mass-Radius relationship in static conditions for pure nucleon star (N), neutron star with hyperons (NH) and hybrid neutron star (HS) with hyperons and quarks with Gibbs construction for two different bag constants. The points of maximum mass obtained are shown. Observational limits imposed from high mass stars PSR J1614-2230 (M = (1.928 $\pm~ 0.017) M_{\odot}$) \cite{Fonseca} and PSR J0348+0432 ($M = 2.01 \pm 0.04 M_{\odot}$) \cite{Ant} are also indicated. The brown horizontal line indicates the canonical mass ($M = 1.4~M_{\odot}$) while mass $M = 1.6~M_{\odot}$ is marked with black. Range of $R_{1.4}$ is marked according to \cite{Abbott,Fattoyev} while that of $R_{1.6}$ is marked according to \cite{Abbott} and \cite{Bauswein}.}
\end{figure}

 The static properties of hybrid stars are calculated with the EoS obtained with Gibbs construction. For $B=(180$ MeV)$^4$, the maximum gravitational mass is obtained as 1.87 $M_{\odot}$ at central density 2.40 $\times 10^{15}$ g cm$^{-3}$ with corresponding radius 11.1 km. For $B=(200$ MeV)$^4$ the maximum gravitational mass is obtained as 1.98 $M_{\odot}$ at central density 1.81 $\times 10^{15}$ g cm$^{-3}$ with corresponding radius 11.6 km. The maximum baryonic mass in each case is found to be 2.19 $M_{\odot}$ and 2.27 $M_{\odot}$, respectively. The values of $R_{1.4}$ and $R_{1.6}$ obtained for $B=(180$ MeV)$^4$ are 12.9 km and 12.4 km, respectively while for $B=(200$ MeV)$^4$ they are 13.3 km and 13.1 km, respectively. In fig. \ref{mr_allG} we show the variation of hybrid star mass with radius for Gibbs construction. Consistent with results of works like \cite{Bhattacharya, Logoteta 2} with different approaches, we find from figs. \ref{mr_all} and \ref{mr_allG}, MC yields more massive configurations of NS due to delayed appearance of quarks compared to that in case of GC. With Gibbs construction the static properties of the hybrid star are tabulated in table \ref{tableGC}
 
\begin{table}[ht!]
\caption{Static neutron star properties for hybrid star (HS) matter for two different bag constants ($B$) with Gibbs construction. The results from hydrostatic equilibrium conditions such as the central density of the star $\varepsilon_c$ ($\times 10^{15}$ g cm$^{-3}$), the maximum gravitational mass $M$ ($M_{\odot}$), maximum baryonic mass $M_B$ ($M_{\odot}$) radius $R$ (km), $R_{1.4}$ (km) and $R_{1.6}$ (km) are displayed}
%{\small{
%\hline
\begin{center}
\begin{tabular}{cccccccccc}
\hline
\hline
\multicolumn{1}{c}{}&
\multicolumn{1}{c}{$B^{1/4}$}&
\multicolumn{1}{c}{$\varepsilon_c$}&
\multicolumn{1}{c}{$M$}&
\multicolumn{1}{c}{$M_{B}$} &
\multicolumn{1}{c}{$R$} &
\multicolumn{1}{c}{$R_{1.4}$} &
\multicolumn{1}{c}{$R_{1.6}$} & \\
%\multicolumn{1}{c}{$M_{B}$} & 
%\multicolumn{1}{c}{$R$} & \\
%
\multicolumn{1}{c}{} &
\multicolumn{1}{c}{(MeV)} &
\multicolumn{1}{c}{($\times 10^{15}$ g cm$^{-3}$)} &
%\multicolumn{1}{c}{} &
%\multicolumn{1}{c}{} &
%\multicolumn{1}{c}{} &
%\multicolumn{1}{c}{$(\rho_0)$} &
\multicolumn{1}{c}{($M_{\odot}$)} &
\multicolumn{1}{c}{($M_{\odot}$)} &
\multicolumn{1}{c}{($km$)} &
\multicolumn{1}{c}{($km$)} &
\multicolumn{1}{c}{($km$)} & \\
\hline
HS &180 &2.40 &1.87 &2.19 &11.1 &12.9 &12.4 \\
HS &200 &1.81 &1.98 &2.27 &12.6 &13.3 &13.1 \\
\hline
\hline
\end{tabular}
\end{center}

\protect\label{tableGC}
\end{table} 

Similar to \cite{Bhattacharya, Logoteta 2} we find that GC yields less massive hybrid stars compared to that with MC. However, the $2.01 \pm 0.04 M_{\odot}$ maximum mass criterion of NS \cite{Ant} is still satisfied in this work with GC. With GC the increase in mass from pure hadronic phase (NH) is (8 - 14)\% whereas it is (10 - 16)\% with MC. There is also feeble drop in central density and baryonic mass with GC compared to that with MC. The radii values of $R_{1.4}$ and $R_{1.6}$ obtained with GC are consistent with the ranges suggested by \cite{Abbott,Fattoyev,Bauswein}.

%\newpage

\end{document}